\documentclass[%
 reprint,
 nolongbibliography,
 amsmath,amssymb,
 aps,
]{revtex4-2}

\usepackage{graphicx}
\usepackage{dcolumn}
\usepackage{bm}
\usepackage{xcolor}

\def\##1{\underline #1}
\def\=#1{\underline{\underline #1}}

\def\H{{\bf H}}

\usepackage{tensor}

\usepackage{hyperref}
\hypersetup{
    colorlinks=true,
    linkcolor=blue,
    citecolor=blue,
    urlcolor=blue,
    breaklinks=true,
}

\begin{document}

\preprint{APS/123-QED}

\title{Coupling Light Waves to Gravitational Waves}

\author{Martin W.\ McCall}
\affiliation{Blackett Laboratory, Department of Physics, Imperial College of Science, Technology and Medicine, Prince Consort Road, London SW7 2AZ, United Kingdom}
\author{Stefanos Fr.\ Koufidis}
\email{steven.koufidis20@imperial.ac.uk}
\affiliation{Blackett Laboratory, Department of Physics, Imperial College of Science, Technology and Medicine, Prince Consort Road, London SW7 2AZ, United Kingdom}

\date{\today}

\begin{abstract}
We demonstrate analytically that gravitational waves, upon interacting with co-propagating electromagnetic radiation {\color{black} in a plasma}, induce distinctive sidebands on the modulated light, thereby providing a detectable signature of their presence. Employing a fully covariant coupled-wave framework, we envision gravitational waves as phase-insensitive ``luminal moving gratings'' and derive explicit phase-matching conditions that articulate such an interaction whilst conserving both energy and momentum. Beyond preserving the directional signature of gravitational waves, the coupling mechanism imposes no coherence requirements on the photon-by-graviton scattering, hence enabling possibilities for exploiting cosmic microwave background radiation. Although detection at low frequencies is constrained by the requirement of long interaction lengths, advances in laser technology are poised to enable high-frequency gravitational wave detection, potentially unveiling insights into the primordial spacetime ripples that have been traversing the cosmos since the inflationary epoch.
\end{abstract}

\maketitle
\par As early as 1893, Heaviside proposed an analogy between electromagnetic waves (EMWs) and gravitational waves (GWs) \cite{heaviside1893}, a concept later formalized through Einstein’s general theory of relativity \cite{einstein1916}. While pioneering efforts in the 1960s laid the groundwork for GW detection \cite{weber1960gravitational}, only recently have advanced interferometric techniques achieved the direct detection of GWs, primarily at low frequencies (10–1000 Hz) \cite{abbot2016observation}. The detection of \emph{high}-frequency GWs in the MHz–GHz range, however, remains an uncharted experimental frontier, with the potential to unveil early-universe phenomena such as cosmic strings \cite{damour2001gravitational}. While resonant microwave cavities \cite{cruise2012potential} and axion haloscope detectors \cite{domcke2022novel} have shown promise, experimental confirmation remains a formidable challenge.

The interaction between EMWs and GWs has captivated physicists since the prediction of the Gertsenshtein effect, whereby an EMW traversing a strong transverse magnetic field generates a GW of the same frequency \cite{gertsenshtein1962wave}, provided both waves are coherent \cite{zeldovich1973electromagnetic}. Despite its theoretical appeal, this effect is limited by weak coupling, which has hindered observational prospects. The inverse Gertsenshtein effect—whereby spacetime distortions within a static magnetic field generate EMWs—further highlights the complexity of such EMWs-GWs interactions \cite{Herman2021detecting}.

\begin{figure}[!t]
\centering
\includegraphics[width=0.46\textwidth]{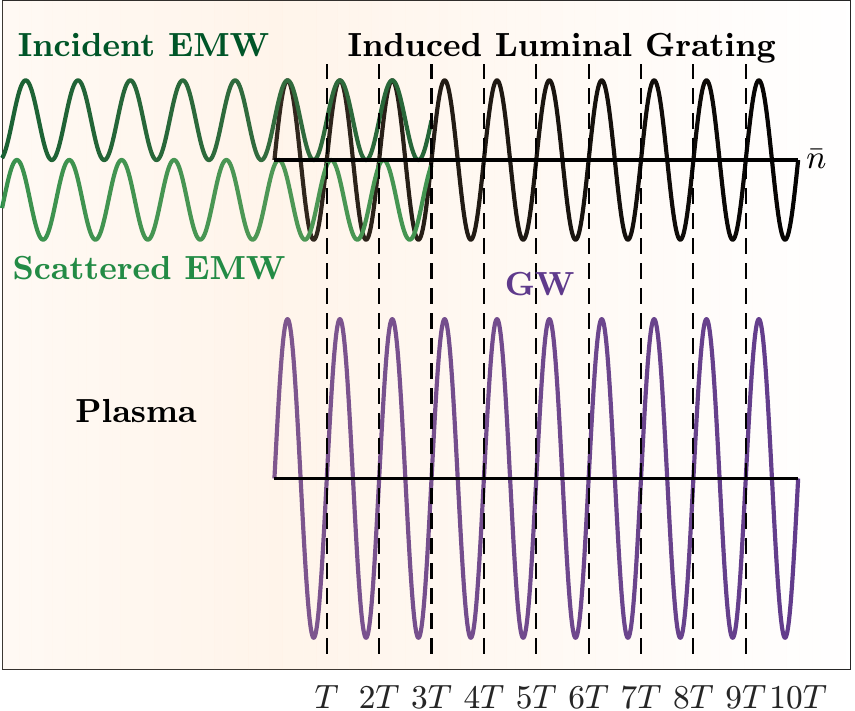}
\caption{\textbf{Luminal Grating Induced by a Gravitational Wave.}
		A plane GW, propagating at the speed of light, induces a periodic (period: $T$) spatiotemporal modulation in the refractive index (mean value: $\bar{n}$), forming a dynamic scaffold that facilitates forward scattering of EMWs {\color{black} in a plasma}. The emerging {\color{black}subluminal} EMWs encode the grating's characteristics, thus enabling indirect GW detection, while preserving its \emph{directional} signature without necessitating \emph{coherence}.}
\label{fig:GraviGrating} 
\end{figure}

Akin to the paradigm of ``synthetic traveling wave modulation'' in meta-optics \cite{pendry2021gain} and the notion of a {\color{black}``luminal mirror'' \cite{esirkepov2024luminal}}, we propose in this communication to conceptualize the latter as ``luminal moving gratings''—see Fig.\ \ref{fig:GraviGrating}. Employing a fully covariant coupled-wave approach, we establish a rigorous framework for EMW-GW interactions, wherein GWs are modeled as archetypal spacetime perturbations propagating at the speed of light. These dynamic disturbances induce effective refractive index modulations, which interact with EMWs {\color{black} in a plasma} under specific phase-matching conditions that conserve both energy and momentum. Our scheme aims to extend the detectable GW frequency spectrum beyond the limitations of current interferometric techniques and provides a pathway towards the long-sought \emph{directional} detection of high-frequency GWs, while relaxing the requirement for \emph{coherence} between the interacting waves.

In the presence of a weak gravitational field, the spacetime metric can be treated as a small perturbation of the flat Minkowski metric $\eta_{\mu\nu}={\rm diag}(-1,1,1,1)$. Thus in nearly Lorentzian coordinates we have $g_{\mu\nu} = \eta_{\mu\nu} + h_{\mu\nu}$, where $h_{\mu\nu}$ is the GW perturbation, with $\left|h_{\mu\nu}\right|$ and all its derivatives $\ll 1$ \cite{supplement}. Furthermore, in the absence of sources, the electromagnetic field in curved spacetime obeys a wave equation for the 4-potential $A^\mu$. In transverse-traceless (TT) gauge the Ricci tensor for {\color{black} non-dispersive} GWs vanishes \footnote{In TT gauge $h = \eta^{\alpha\beta} h_{\alpha\beta} = h\indices{_\alpha^\alpha} = 0$. Substituting the  GW ansatz into the Ricci tensor yields: $2R_{\mu\nu} = -K_\mu K_\alpha h\indices{_\nu^\alpha} - K_\nu K_\alpha h\indices{_\mu^\alpha} + K_\alpha K^\alpha h\indices{_{\mu\nu}} - h\indices{_\alpha^\alpha_{,\mu\nu}}$. The first two terms vanish on account of transversality, the third term vanishes since $K_\alpha$ is null, and the fourth term vanishes due to tracelessness; $K_\mu$ is the GW's 4-wavevector.} and, under the Lorenz gauge, $A\indices{^{\mu}_{;\mu}}=0$, the wave equation simplifies to $A\indices{^{\alpha;\mu}_{;\mu}}=0$ \footnote{The TT gauge for weak GWs simplifies the Lorenz gauge to $A\indices{^{\mu}_{,\mu}} = 0$, ensuring compatibility: the former fixes the coordinates for $h_{\mu\nu}$, the latter is metric-independent.}. In linearized gravity the latter is developed as \footnote{In Lorenz gauge the wave equation is $A\indices{^{\alpha;\mu}_{;\mu}}=R\indices{^\alpha_{\mu}}A^\mu$, which for vanishing Ricci tensor reduces to $A\indices{^{\alpha;\mu}_{;\mu}}=0$. The covariant derivatives are expanded as $A\indices{^{\alpha,\mu}_{,\mu}}+\eta^{\mu\lambda}\Gamma\indices{^\alpha_{\nu\lambda,\mu}}A^\nu
+\Gamma\indices{^\alpha_{\nu\lambda}}A^{\nu,\lambda}
+\left(A^{\beta,\mu}+\eta^{\mu\lambda}\Gamma\indices{^\beta_{\nu\lambda}}A^\nu\right)\Gamma\indices{^\alpha_{\beta\mu}}=0$. As detailed in Sec.\ V of the Supplement \cite{supplement}, the Christoffel symbols in the linearized regime retain only $\mathcal{O}\left(h\right)$ terms, thereby leading to Eq.\ \eqref{eq:WaveEquationFinal}.}
\begin{equation}\label{eq:WaveEquationFinal}
    A\indices{^{\alpha,\mu}_{,\mu}} + \eta^{\mu\lambda} \Gamma\indices{^\alpha_{\nu\lambda,\mu}} A^\nu + 2 \Gamma\indices{^\alpha_{\nu\lambda}} A^{\nu,\lambda} = 0\,,
\end{equation}
where $\Gamma\indices{^\alpha_{\nu\lambda}}$ are the Christoffel symbols and the Einstein summation convention is herein implicit.

For a metric perturbation of the form $h^{\mu\nu}=\left({1}/{2}\right)\mathcal{H}^{\mu\nu}e^{iK_\rho x^\rho}+\text{c.c.}$, $\mathcal{H}^{\mu\nu}$ is a Hermitian polarization tensor, $K_\mu =\Omega(-1,0,0,\pm1)$ is the 4-wavevector of the GW, and complex conjugation ensures real solutions; $\Omega$ is the GW frequency and the ``$\pm$'' sign indicates the direction of phase propagation. 

Considering first-order diffracted sidebands, the total EMW potential is formulated as a superposition of an incident, ${\cal A }_{(0)}$, and two scattered, ${\cal A }_{(-1)}$ and ${\cal A }_{(1)}$, wave potentials: $A^\alpha =\sum_j{\cal A }_{(j)}^\alpha  e^{i k_{(j)\mu} x^\mu}$. For a plane EMW propagating {\color{black} in a plasma} along the $z$-axis, {\color{black}$k_{{(j)}\mu} = \omega_{(j)}(-1,0,0,\pm n)$} is the 4-wavevector, with $\omega_{(j)}$ being both the temporal and the spatial frequency in units where the phase velocity of light in vacuum is unity; $j = \{-1,0,1\}$ labels the incident, $j=0$, and scattered, $j=\pm 1$, EMWs. {\color{black} In a plasma, dispersion modifies the refractive index, which in the high-frequency limit may be approximated as $n(\omega) = 1 - \delta n$, where $0 < \delta n \ll 1$. This slight departure from lightlike propagation nullifies the vacuum null condition: $k_{(j)}^\mu k_{(j)\mu} = \omega_{(j)}^2(n^2 - 1) \approx -2\delta n\omega_{(j)}^2$. In typical interstellar environments, $\delta n \sim 10^{-8}$ at radio frequencies \cite{Draine2011}—a small yet consequential deviation permitting interactions with GWs otherwise forbidden in the luminal regime \cite{mahajan2023parametric, asenjo2024upshifted}.}

\par Substituting the 4-potential ansatz into Eq.\ \eqref{eq:WaveEquationFinal}, the slowly varying envelope approximation leads to \footnote{Terms of Eq.\ \eqref{eq:WaveEquationFinal}: $A^{\nu,\lambda} = \sum_j \left( {\cal A }_{(j)}^{\nu,\lambda} + i k_{(j)}^\lambda {\cal A }_{(j)}^\nu \right) e^{i k_{(j)\mu} x^\mu}$ and $A\indices{^{\alpha,\mu}_{,\mu}} \approx \sum_j \left(  2ik_{(j)}^{\mu} {\cal A }^{\alpha}_{(j),\mu}{\color{black}
- k_{(j)}^\mu k_{(j)\mu}{\cal A }_{(j)}^\alpha} \right) e^{ik_{(j)\nu}x^\nu}$. {\color{black} In a plasma, introducing a constitutive tensor $\chi\indices{^{\alpha\beta\mu\nu}}$, the mode condition $\chi\indices{^{\alpha\beta\mu\nu}}k_\mu k_\beta = 0$ of Eq.\ (6.35) in Ref.\ \cite{Post1997} eliminates the rightmost term in $A\indices{^{\alpha,\mu}_{,\mu}}$.}}
\begin{widetext}
    \begin{align}\label{eq:cwtlong}
             &\sum_j\left[2i k_{(j)}^{\mu} {\cal A }^{\alpha}_{(j),\mu} +i\frac{\eta^{\alpha\beta}}{2}\left(\Delta\Gamma\indices{_{\beta\nu\lambda}}e^{iK_\rho x^\rho} - \Delta\Gamma\indices{^{*}_{\beta\nu\lambda}}e^{-iK_\rho x^\rho} \right){\cal A }_{(j)}^{\nu,\lambda}\right]e^{i k_{(j)\gamma}x^\gamma} =  \nonumber
        \\
        &\sum_j\left[\frac{\eta^{\alpha\beta}}{2}\left(\Delta\Gamma\indices{_{\beta\nu\lambda}}e^{iK_\rho x^\rho} -\Delta\Gamma\indices{^{*}_{\beta\nu\lambda}}e^{-iK_\rho x^\rho}\right)k_{(j)}^\lambda-\frac{\eta^{\mu\lambda}\eta^{\alpha\beta}}{4}\left(\Delta\Gamma\indices{_{\beta\nu\lambda\mu}}e^{iK_\rho x^\rho}+\Delta\Gamma\indices{^{*}_{\beta\nu\lambda\mu}} e^{-iK_\rho x^\rho}\right)  \right]{\cal A }_{(j)}^\nu e^{ik_{(j)\mu} x^\mu}\,,
    \end{align}
\end{widetext}
where $\Delta\Gamma\indices{_{\beta\nu\lambda}}=\mathcal{H}_{\beta\lambda}K_\nu+\mathcal{H}_{\nu\beta}K_\lambda-\mathcal{H}_{\nu\lambda}K_\beta$ and $ \Delta\Gamma\indices{_{\beta\nu\lambda\mu}}= \mathcal{H}_{\nu\lambda}K_\beta K_\mu-\mathcal{H}_{\beta\lambda}K_\nu K_\mu-\mathcal{H}_{\nu\beta}K_\lambda K_\mu$.

Scrutinizing Eq.\ \eqref{eq:cwtlong} one can forthwith identify that synchronous sidebands must have 4-wavevectors given by
\begin{equation}\label{eq:phase-match}
        k_{(1)\mu}=k_{(0)\mu} + K_\mu \quad {\rm and} \quad k_{(-1)\mu}=k_{(0)\mu}-K_\mu\,,
\end{equation}
hence reaffirming the simulations of Ref.\ \cite{falcon2023interactions}, and consistent with 4-momentum conservation  required for single-graviton absorption in the dilute regime of Ref.\ \cite{Carney2024signature}. The phase-matching conditions of Eq.\ \eqref{eq:phase-match} encapsulate the requirement that for this {\color{black}subluminal \footnote{{\color{black} 
Above the plasma frequency $\omega_p$, EMWs in a plasma are regarded as subluminal because their group velocity, $v_{\mathrm{g}} = \left(1 - \omega_p^2 / \omega_{(0)}^2\right)^{1/2}$, falls below unity, even though their phase velocity, $v_{\mathrm{ph}} = \left(1 - \omega_p^2 / \omega_{(0)}^2\right)^{-1/2}$, exceeds it.}}} coupling process both \emph{energy} ($\mu = 0$) and \emph{momentum} ($\mu = 3$) must be simultaneously conserved, dictating moreover that {\color{black}for weak $\delta n$} \emph{all} EMWs and GWs must be \emph{co}-propagating. It is this aspect of the coupling that conveys directional information about the GWs, with the angular spectrum of the generated sidebands being, in practice, dictated by the bandwidths of both the GW and the primary EMW.

\par Manifestly, the relationships in Eq.\ \eqref{eq:phase-match} represent an intermediate regime between a static Bragg grating, which conserves energy but not momentum \cite{koufidis2022mobius}, and a temporal Bragg grating, which conserves momentum but not energy \cite{koufidis2023temporal}. This intermediary scenario pertains to traveling-wave modulation \cite{pendry2021gain, horsley2024traveling}, with Eq.\ \eqref{eq:phase-match} aligning with the frame-specific conservation laws derived in Ref.\ \cite{Zhang2024conservation}. A key distinction from the preceding cases lies in the absence of a resonant (or indeed synchronization) phenomenon \emph{per se}, evincing that the ``scattering'' effect is \emph{not} contingent upon any specific resonant frequency.

\par Given the heretofore prevailing technological limitations for GW detection, we assume $\left|K_\mu\right| \ll \left|k_{(j)\mu}\right|$, indicating that light couples with a GW of considerably lower frequency. Under this assumption, $\omega_{(0)} \approx  \omega_{(1)} \approx \omega_{(-1)} \equiv \omega$, substantiating that the outgoing wave phases co-propagate with respect to the incident wave phase. Nevertheless GWs are either predicted or observed across a wide spectral range. Therefore, by relaxing this presupposition and setting, say, $\Omega = 2\omega_{(0)}$, Eq.\ \eqref{eq:phase-match} entails that the spatial and temporal directions of the ``$(-1)$'' generated sideband are reversed from {\color{black}$k_{(-1)\mu} = \omega_{(0)}(-1,0,0,+n)$ to $k_{(-1)\mu} = \omega_{(0)}(1,0,0,-1-\delta n)$}. Since EMWs are described by the real part of phasors, the ``$(-1)$'' generated sideband will remain co-propagating with respect to the incident wave, underpinning yet another fundamental distinction from conventional Bragg scattering: {\color{black} for most natural environments, where $\delta n$ is weak,} no backscattering as such is supported  \footnote{{\color{black}For co-propagating waves in a plasma, backscattering of a frequency downshifted ``$(-1)$'' EMW requires $\delta n > 1 - \Omega / \omega_{(0)}$, provided that $\omega_{(0)} > \Omega$. Conversely, for counter-propagating waves in a plasma, backscattering of a frequency upshifted ``$(+1)$'' EMW requires the same inequality, which in this instance holds for arbitrary values of $\omega_{(0)}$ and $\Omega$. One might hastily conclude that setting $\delta n = 0$ always permits backscattering of counter-propagating EMWs and GWs; however, this is incorrect, as in the luminal regime the $k_0$ and $k_3$ components of the 4-wavevector must be equal in magnitude.}}.

\par Applying the phase-matching conditions from Eq.\ \eqref{eq:phase-match} we can model the interaction between the impinging EMW and the forward-scattered EMWs {\color{black} in a plasma}, mediated by the GW, via a system of coupled-wave equations. Collecting all terms that match $k_{(0)}$ we obtain
\begin{subequations}
\begin{align}
    k_{(0)}^{\mu} {\cal A }^{\alpha}_{(0),\mu} +C\indices{^\alpha_\nu^\mu}{\cal A }^{\nu}_{(-1),\mu} - \left(C\indices{^\alpha_\nu^\mu}\right)^*{\cal A }^{\nu}_{(1),\mu} \nonumber
    \\
    = P\indices{^\alpha_\nu}{\cal A}^\nu_{(-1)}  +Q\indices{^\alpha_\nu}{\cal A}^\nu_{(1)}\,. \label{eq:cwe A0compact}
\end{align}
Subsequently the terms that match $k_{(1)}$ are written as
\begin{align}
    k^\mu_{(1)}{\cal A}^\alpha_{(1),\mu} + C\indices{^\alpha_\nu^\mu}{\cal A}^\nu_{(0),\mu} =M\indices{^\alpha_\nu}{\cal A}^\nu_{(0)}+iN\indices{^\alpha_\nu}{\cal A}^\nu_{(0)}\,,
                \label{eq:cwe A1compact}
\end{align}
while those that match $k_{(-1)}$ as
\begin{align}
k^\mu_{(-1)}{\cal A}^\alpha_{(-1),\mu} - \left(C\indices{^\alpha_\nu^\mu}\right)^*{\cal A}^\nu_{(0),\mu} \nonumber
    \\
    = \left(M\indices{^\alpha_\nu}\right)^*{\cal A}^\nu_{(0)} +i \left(N\indices{^\alpha_\nu}\right)^*{\cal A}^\nu_{(0)}\,.
                \label{eq:cwe Am1compact}  
\end{align}
\end{subequations}
The introduced tensors are defined as follows: $C\indices{^{\alpha}_{\nu}^{\mu}} = \left({\eta^{\alpha\beta}}/{4}\right) \Delta\Gamma\indices{_{\beta\nu\lambda}}\eta^{\lambda\mu}$, $M\indices{^\alpha_\nu} = -i\left({\eta^{\alpha\beta}}/{4}\right)k_{(0)}^\lambda\Delta\Gamma\indices{_{\beta\nu\lambda}}$, $N\indices{^\alpha_\nu} = \left({\eta^{\mu\lambda}\eta^{\alpha\beta}}/{8}\right)\Delta\Gamma_{\beta\nu\lambda\mu}$, $Q\indices{^\alpha_\nu} = \left({i}{/4}\right)\left[\eta^{\alpha\beta}k_{(1)}^\lambda\Delta\Gamma\indices{^*_{\beta\nu\lambda}}+\left({\eta^{\mu\lambda}\eta^{\alpha\beta}}/{2}\right) \Delta\Gamma\indices{^*_{\beta\nu\lambda\mu}}\right]$, and $ P\indices{^\alpha_\nu} = -\left({i}/{4}\right)\left[\eta^{\alpha\beta}k^\lambda_{(-1)}\Delta\Gamma_{\beta\nu\lambda}-\left({\eta^{\mu\lambda}\eta^{\alpha\beta}}/{2}\right)\Delta\Gamma_{\beta\nu\lambda\mu}\right]$.

Considering transverse waves {\color{black}traversing a plasma} along the $z$-axis we may define $
\mathbf{A} = \begin{bmatrix}
{\cal A}^x_{(-1)} \
{\cal A}^y_{(-1)} \
{\cal A}^x_{(0)} \
{\cal A}^y_{(0)} \
{\cal A}^x_{(1)} \
{\cal A}^y_{(1)}
\end{bmatrix}^\intercal
$, where $^\intercal$ denotes transpose, and calculate the non-zero independent components of the aforementioned tensors \footnote{Non-zero components per tensor follow. \underline{Tensor $ C\indices{^{\alpha}_{\nu}^{\mu}}$}: $C\indices{^{0}_{1}^{1}}$, 
    $C\indices{^{0}_{1}^{2}}$, $C\indices{^{0}_{2}^{1}}$, 
    $C\indices{^{0}_{2}^{2}}$, 
    $C\indices{^{1}_{0}^{1}}$, 
    $C\indices{^{1}_{1}^{0}}$, 
    $C\indices{^{1}_{0}^{2}}$, 
    $C\indices{^{1}_{2}^{0}}$,  
    $C\indices{^{1}_{1}^{3}}$, 
    $C\indices{^{1}_{3}^{1}}$, 
    $C\indices{^{1}_{2}^{3}}$, 
    $C\indices{^{1}_{3}^{2}}$, 
    $C\indices{^{2}_{0}^{1}}$, 
    $C\indices{^{2}_{1}^{0}}$, 
    $C\indices{^{2}_{0}^{2}}$, 
    $C\indices{^{2}_{2}^{0}}$,  
    $C\indices{^{2}_{1}^{3}}$, 
    $C\indices{^{2}_{3}^{1}}$, 
    $C\indices{^{2}_{2}^{3}}$, 
    $C\indices{^{2}_{3}^{2}}$, 
    $C\indices{^{3}_{1}^{1}}$, 
    $C\indices{^{3}_{1}^{2}}$, 
    $C\indices{^{3}_{2}^{1}}$, 
    $C\indices{^{3}_{2}^{2}}$. \underline{Tensor $M\indices{^\alpha_\nu}$}: $M\indices{^1_1}$, $M\indices{^1_2}$, $M\indices{^2_1}$, $M\indices{^2_2}$. \underline{Tensor $Q\indices{^\alpha_\nu}$}: $Q\indices{^1_1}$, $Q\indices{^1_2}$, $Q\indices{^2_1}$,  $Q\indices{^2_2}$. \underline{Tensor  $P\indices{^\alpha_\nu}$}: $P\indices{^1_1}$, $P\indices{^1_2}$, $P\indices{^2_1}$,  $P\indices{^2_2}$. \underline{Tensor $N\indices{^\alpha_\nu}$}: $\emptyset$. For analytic expressions see Sec.\ VII of the Supplement \cite{supplement}.}. Thence, by setting $\alpha\equiv x$ and $\alpha\equiv y$ we can concisely cast Eqs.\ \eqref{eq:cwe A0compact}, \eqref{eq:cwe A1compact}, and \eqref{eq:cwe Am1compact} in a matrix notation, viz.,
\begin{equation}\label{CWEs General}
    \mathbf{L} \cdot {\rm d}_u \mathbf{A} = \mathbf{S} \cdot \mathbf{A}\,,
\end{equation}
where the operator ${\rm d}_u = {\partial}/{\partial t} + {\color{black}n}{\partial}/{\partial z}$ differentiates along the {\color{black}\emph{timelike}} curve tracking the EMWs; $u$ is an affine parameter. Additionally we have
\begin{equation*}
    \mathbf{L} =
    \begin{bmatrix}
        \omega_{(-1)} \mathbb{I} & \frac{\Omega}{4} \mathbf{S}_{+,\times}^* & \mathbf{0} \\[10pt]
        -\frac{\Omega}{4} \mathbf{S}_{+,\times} & \omega_{(0)} \mathbb{I} & \frac{\Omega}{4} \mathbf{S}_{+,\times}^* \\[10pt]
        \mathbf{0} & \frac{\Omega}{4} \mathbf{S}_{+,\times} & \omega_{(1)} \mathbb{I}
    \end{bmatrix} \,
\end{equation*}
and
\begin{equation*}
    \mathbf{S} = i{\color{black}\delta n}\frac{\Omega}{4}
    \begin{bmatrix}
       \mathbf{0} & -\omega_{(0)}
        \mathbf{S}_{+,\times}^* & \mathbf{0} \\[10pt]
        \omega_{(-1)}
         \mathbf{S}_{+,\times} & \mathbf{0} &
        -\omega_{(1)}
        \mathbf{S}_{+,\times}^* \\[10pt]
        \mathbf{0} & \omega_{(0)}
         \mathbf{S}_{+,\times} & \mathbf{0}
    \end{bmatrix} \,;
\end{equation*}
$\mathbb{I}$ is the $2\times2$ identity, $\mathbf{0}$ is the $2\times2$ null matrix, and
\begin{equation*}
    \mathbf{S}_{+,\times}=\begin{bmatrix}
            \mathcal{H}_+ & \mathcal{H}_\times \\
            \mathcal{H}_\times & -\mathcal{H}_+
        \end{bmatrix}\,,
\end{equation*}
with $\mathcal{H}_+$ and $\mathcal{H}_\times$ being the amplitudes of the ``plus'' and ``cross'' polarization states of the GW, respectively \cite{supplement}.

Insofar as $\mathbf{L}$ is invertible, we may set $\mathbf{W} = \mathbf{L}^{-1} \cdot \mathbf{S}$ so that the coupled-wave system in Eq.\ \eqref{CWEs General} becomes
\begin{equation}\label{eq:4dcwequation}
  {\rm d}_u \mathbf{A} =  \mathbf{W}\left({\color{black}\delta n}, \omega_{(0)},\Omega,\mathcal{H}^{\mu\nu}\right) \cdot \mathbf{A}\,.
\end{equation}
Although the linear system in Eq.\ \eqref{eq:4dcwequation} is autonomous and thus admits analytic closed-form solutions, its foremost merit resides in yielding straightforward insights. Indeed, for the specified polarization, we could approximate the coupling strengths between the impinging and the forward-scattered EMWs as $\kappa_{(\pm 1)}\approx {\color{black}\delta n}\left[{\omega_{(0)}\Omega}/{\left({\color{black}4}\omega_{(\pm 1)}\right)}\right]h_s$; $h_s=\left|\mathcal{H}_{+}\right|$ or $\left|\mathcal{H}_\times\right|$ \footnote{The diagonal terms of $\mathbf{L}$ vastly outweigh the $\mathcal{O}(h_s)$ off-diagonal terms. Whence, e.g., the $(3,2)$ element of $\mathbf{W}$ is $\kappa^{x,y}_{-1,0} ={\color{black}\delta n} {\omega_{(0)}\Omega\mathcal{H}_\times^*}/{\left(2\omega_{(-1)}\right)}$—cf.\ Eq.\ \eqref{CWEsReduced} in the Supplement \cite{supplement} and the coupling coefficients therein}. Whence the relative sidebands amplitudes are $\left|{{\cal A }_{(\pm 1)}}/{{\cal A }_{(0)}}\right| \approx \left({4}/{\pi}\right)\kappa_{(\pm 1)} L_{c,(\pm 1)}$, with $L_{c,(\pm 1)}$ being the characteristic interaction length of each diffraction order \cite{yariv1973coupled}. Translating this into an observable intensity ratio, $\mathcal{I}_{(j)}\approx\left|\mathcal{A}_{(j)}\right|^2$, just above the noise threshold relative to the primary laser intensity, we estimate the interaction lengths as
\begin{equation}\label{eq:estimation}
    L_{c, (\pm 1)} \approx {\color{black}\frac{1}{2\delta n}}\frac{\lambda_{(0)}}{\lambda_{(\pm 1)} }\frac{\Lambda}{h_s}\sqrt{\frac{\mathcal{I}_{(\pm 1)}}{\mathcal{I}_{(0)}}}\,.
\end{equation}
Here $\Lambda = {2\pi}/{\Omega}$ and $\lambda_{\{(0), (\pm 1)\}} = {2\pi}/{\omega_{\{(0), (\pm 1)\}}}$ are the wavelengths of the GW and the EMWs, respectively.

\begin{figure}
	\centering
	\includegraphics[width=0.46\textwidth]{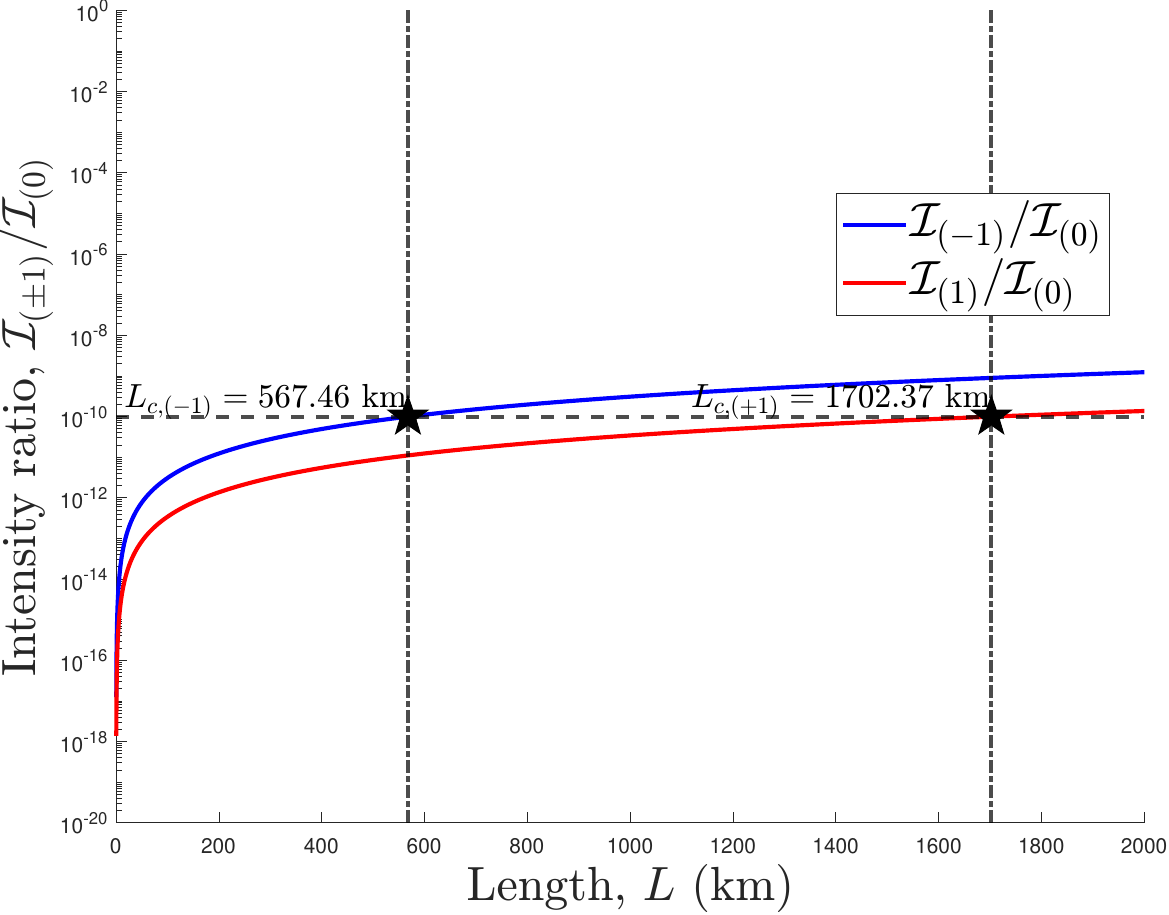} 
\caption{\textbf{Intensity Ratio vs.\ {\color{black} Propagation} Length.}
The intensity ratios, ${\mathcal{I}_{(\pm 1)}}/{\mathcal{I}_{(0)}}$, are plotted as functions of the corresponding {\color{black} propagation length $L$} on a semi-logarithmic scale. The solid lines represent the calculated intensity ratios based on numerical integration of Eq.\ \eqref{eq:4dcwequation}, while the dashed line shows the noise threshold, $\left[{\mathcal{I}_{(\pm 1)}}/{\mathcal{I}_{(0)}}\right]_{\rm thresh}=10^{-10}$, taken from the intensity noise spectrum of the ultra-low-noise solid-state laser proposed in Ref.\ \cite{shoji2016ultra}. {\color{black}Within dense layers of the stellar atmosphere we take $\delta n=0.14$,} and the parameters achievable with the next-generation National Ignition Facility are: $\lambda_{\rm (0)} = 352.94$ nm, $\Lambda = 176.47$ nm, and $h_s = 10^{-17}$ {\color{black}(cf.\ Fig.~2 in Ref.~\cite{vacalis2023detection}, wherein resonant enhancement enables sensitivity to such strains without violating energy conservation). Modulo strain-enhancement considerations, the condition $\Omega = 2\omega_{(0)}$ is merely illustrative and \emph{not} required, unlike in cases of parametrically amplified GWs \cite{mahajan2023parametric}.} N.B.—In this {\color{black} nearly-luminal} scenario we approximate the affine parameter $u$ in the quasi-static limit as $ u \equiv {\color{black}n}z \equiv t $.}
\label{fig:IntensityRatioVsLengthTerr}
	
\end{figure}

{\color{black} 

\par Since the identified mechanism does not rely on phase coherence, incoherent sources, such as the cosmic microwave background (30--600~GHz), become viable probes of high-frequency GWs. At the indicative wavelength of $\lambda_{(0)} = 2 \times 10^{-3}\,\mathrm{m}$ propagation requires $\omega_{(0)} > \omega_p$, where $\omega_p$ is the plasma frequency—a condition satisfied in galactic disk regions with electron densities $n_e \sim 10^{-2}\,\mathrm{cm}^{-3}$ yielding a dispersion parameter $\delta n \sim 10^{-17}$ \cite{Draine2011}. For currently detectable GWs we adopt $\Lambda = 10^3\,\mathrm{m}$ and $h_s = 10^{-21}$; then, assuming $\left[{\mathcal{I}_{(\pm 1)}}/{\mathcal{I}_{(0)}}\right]_{\rm thresh} = 10^{-10}$, Eq.~\eqref{eq:estimation} yields $L_{c,(\pm 1)} \sim 10^{35}\,\mathrm{m}$ for both sidebands. This distance, though finite, is alas \emph{too} large, even by cosmological standards (the diameter of the observable universe is $D_{\mathrm{U}} \approx 8.8 \times 10^{26}\,\mathrm{m}$). A coupled-wave theory estimate of the interaction lengths required for the cosmic microwave background radiation to generate sidebands that could, in principle, be detected with current technology is displayed in Tab.\ \ref{Table_Interaction_Lenghts}.

\begin{table}[h!]
\color{black}
\centering
\renewcommand{\arraystretch}{1.3}
\begin{tabular}{|c|c|c|c|}
\hline
\textbf{Environment} & $n_e\;(\mathrm{cm}^{-3})$ & $\sim\delta n$ & $\sim L_c\;(\mathrm{m})$ \\
\hline
Intergalactic space & $< 10^{-4}$ & $<10^{-19}$& $>10^{37}$ \\
Galactic disk       & $10^{-2}$      & $10^{-17}$& $10^{35}$ \\
H\,\textsc{ii} region & $10^{4}$    & $10^{-11}$ & $10^{29}$ \\
Stellar wind        & $10^{8}$    & $10^{-7}$ & $10^{25}$ \\
\hline
\end{tabular}
\caption{Order-of-magnitude estimates of the interaction lengths required in various astrophysical environments to produce detectable sideband levels. A representative wavelength of $\lambda_{(0)} = 2 \times 10^{-3}\, \mathrm{m}$ is assumed for cosmic microwave background radiation. Evidently, nearly all interaction lengths far exceed the diameter of the observable universe.
}
\label{Table_Interaction_Lenghts}	
\end{table}

\par At optical frequencies, $\omega_{(0)} = 17\pi \times 10^{14}\,\mathrm{Hz}$, the same mechanism—now acting in \emph{extremely} dense stellar atmospheres with $n_e \sim 10^{21}\,\mathrm{cm}^{-3}$—yields an augmented dispersion parameter $\delta n = 0.14$. Assuming the presence of a high-frequency GW with $\Omega = 2\omega_{(0)}$ and $h_s = 10^{-17}$ (cf.\ Fig.~2 in Ref.~\cite{vacalis2023detection}, where strain sensitivity benefits from a resonant enhancement factor $\omega_{(0)}^2 \tau^2$, with $\tau$ being the interaction time), we obtain $\lambda_{(0)} = 352.94\,\mathrm{nm}$ and $\Lambda = 176.47\,\mathrm{nm}$. These yield interaction lengths $L_{c,(-1)} \approx 1.89 \times 10^6\,\mathrm{m}$ and $L_{c,(+1)} \approx 6.29 \times 10^5\,\mathrm{m}$ for the ``$(-1)$'' and ``$(+1)$'' sidebands, respectively. Although considerable, these lengths could be reduced through resonant enhancement, the use of optical cavities (with a plasma generating a strong $\delta n$), or multi-pass interferometric techniques, thus enabling \emph{terrestrial} detection (for comparison, Earth’s diameter is $D_{\oplus} \approx 1.27 \times 10^7\,\mathrm{m}$).
} 

\par The estimated interaction lengths closely align with the computed values determined via brute-force numerical integration of Eq.\ \eqref{eq:4dcwequation}. In fact, the exact values are indicated by the starred points in Fig.\ \ref{fig:IntensityRatioVsLengthTerr}, demonstrating their consistency with current interferometric capabilities and corroborating our estimate. The exponential amplification-like behavior exhibited by the curves in Fig.\ \ref{fig:IntensityRatioVsLengthTerr} accords with the ``compression of lines of force'' gain mechanism of synthetic luminal gratings \cite{pendry2021gain}.

In conclusion, interactions between co-propagating {\color{black}subluminal} EMWs {\color{black} in a plasma} and {\color{black}luminal} GWs within a fully covariant coupled-wave framework reveal that phase-matching conditions—conserving both energy and momentum—predict observable sidebands in the electromagnetic spectrum \cite{falcon2023interactions}, thereby providing a direction-preserving, phase-insensitive method for GW detection. Harnessing static modulations achieved via synthetic gratings \cite{pendry2021gain, horsley2024traveling, harwood2024superluminal} could enable an all-optical analog experiment in laboratory settings. The phase-insensitive mechanism discussed herewith offers a {\color{black}potential} avenue for exploiting incoherent electromagnetic sources, such as the cosmic microwave background, wherein B-mode patterns are hypothesized to encode imprints of primordial GWs \cite{seljak1997signature}. Investigation of stochastic waves is essential to overcoming resolution constraints imposed by the intrinsic variance of the uniform background \cite{mentasti2023intrinsic} and should serve as a cornerstone for future endeavors.

\begin{acknowledgments}
The authors are supported by a Kickstarter Research Project funded by the Department of Physics at Imperial College London. S. F. K. thanks the Bodossaki Foundation for its continued financial support since 2021.  Fruitful discussions with C. R. Contaldi, J. W. G. Tisch, R. Sapienza, G. Mentasti, and Z. Hayran are acknowledged. Mistakes are ours.
\end{acknowledgments}

\bibliography{sorsamp.bib}

\clearpage
\onecolumngrid 

\begin{center}
    {\Large \textbf{Supplemental Material: \\[0.2cm]
    Coupling Light Waves to Gravitational Waves}} \\[0.4cm]
    {\large Martin W.\ McCall and Stefanos Fr.\ Koufidis} \\[0.1cm]
    \textit{Blackett Laboratory, Department of Physics, Imperial College of Science, Technology and Medicine, Prince Consort Road, London SW7 2AZ, United Kingdom} \\[0.1cm]
    (Dated: \today)
\end{center}

\setcounter{equation}{0}
\setcounter{figure}{0}
\setcounter{table}{0}
\setcounter{section}{0}
\setcounter{page}{1}

\renewcommand{\theequation}{S\arabic{equation}}

\makeatletter
\renewcommand{\section}{\@startsection{section}{1}{0pt}%
    {-0.5ex plus -0.2ex minus -.2ex}%
    {0.5ex plus .2ex}%
    {\normalfont\small\bfseries}}
\makeatother

\vspace{0.4cm}
\section{Linearized Theory of Gravity}
\label{Supp:Linearized Gravity}
\vspace{0.4cm}
We begin by assuming that spacetime is only slightly perturbed from flat Minkowski geometry so that the metric of the slightly deformed manifold is given by 
\begin{equation}\label{eq:PerturbedMetric}
g_{\mu\nu} = \eta_{\mu\nu} + h_{\mu\nu} + \mathcal{O}(h)\,, 
\end{equation}
where $h = \eta^{\alpha\beta} h_{\alpha\beta}$ is the trace of $h_{\mu\nu}$. The inverse metric, accurate to first order in $h_{\mu\nu}$, is $g^{\mu\nu} = \eta^{\mu\nu} - h^{\mu\nu}$, with $h^{\mu\nu} = \eta^{\mu\alpha} \eta^{\nu\beta} h_{\alpha\beta}$.

Thence the Christoffel symbols, $\Gamma\indices{^\mu_{\alpha\beta}} = \left({1}/{2}\right) g^{\mu\rho} \left( g_{\rho\beta,\alpha} +  g_{\rho\alpha,\beta} -  g_{\alpha\beta,\rho} \right)$, can be approximated by
\begin{equation}\label{eq:LinerChristoffelSymbols}
\Gamma\indices{^\alpha_{\nu\lambda}}\approx\frac{1}{2}\eta^{\alpha\beta}(h_{\beta\lambda,\nu}+h_{\nu\beta,\lambda}-h_{\nu\lambda,\beta})\,.
\end{equation}
Wherefore quadratic terms in the full expression of the Riemann curvature tensor can be neglected, thus bringing the latter into the approximate form
\begin{equation}\label{eq:LinearRiemann}
R_{\alpha\mu\beta\nu} \approx \Gamma\indices{_{\alpha\mu\nu,\beta}} - \Gamma\indices{_{\alpha\mu\beta,\nu}} = \frac{1}{2} \left( h_{\alpha\nu,\mu\beta} + h_{\mu\beta,\nu\alpha} - h_{\mu\nu,\alpha\beta} - h_{\alpha\beta,\mu\nu} \right).
\end{equation}
Upon contracting the Riemann tensor we obtain the Ricci tensor
\begin{equation}\label{eq:LinearRicciTensor}
    R_{\mu\nu} = R^\alpha_{\mu\alpha\nu} \approx \frac{1}{2} \left( h\indices{_\nu^\alpha_{,\mu\alpha}} + h\indices{_\mu^\alpha_{,\nu\alpha}} - h\indices{_{\mu\nu,}_{\alpha}^\alpha} - h_{,\mu\nu} \right)\,.
\end{equation}
Finally, taking the trace of the Ricci tensor yields the Ricci scalar which at linear order becomes
\begin{equation}\label{eq:LinearRicciScalar}
    R = g^{\mu\nu} R_{\mu\nu} \approx h\indices{^{\mu\alpha}_{,\mu\alpha}} - h\indices{_{,\alpha}^{\alpha}}\,.
\end{equation}

If $G_{\mu\nu} = R_{\mu\nu} - \left({1}/{2}\right) g_{\mu\nu} R$ is the Einstein tensor, setting $g_{\mu\nu} \approx \eta_{\mu\nu}$ and substituting $R_{\mu\nu}$ and $R$ from Eqs.\ \eqref{eq:LinearRicciTensor} and \eqref{eq:LinearRicciScalar}, respectively, yields
\begin{equation}\label{eq:LinearEinsteinEquation}
    G_{\mu\nu} = \frac{1}{2} \left[ h\indices{_{\mu\alpha,\nu}^{\alpha}} + h\indices{_{\nu\alpha,\mu}^{\alpha}} - h\indices{_{\mu\nu,\alpha}^{\alpha}} - h_{,\mu\nu}
    - \eta_{\mu\nu} \left( h\indices{_{\alpha\beta,}^{\alpha\beta}} - h\indices{_{,\beta}^{\beta}} \right) \right].
\end{equation}
To further simplify Eq.\ \eqref{eq:LinearEinsteinEquation} we introduce the trace-reversed metric perturbation \cite{Misner1973}
\begin{equation*}
    \bar{h}_{\mu\nu} = h_{\mu\nu} - \frac{1}{2} \eta_{\mu\nu} h\,,
\end{equation*}
so that Eq.\ \eqref{eq:LinearEinsteinEquation} reduces to the field equation
\begin{equation}\label{eq:EinsteinLinear}
    G_{\mu\nu} = \frac{1}{2} \left( - {\bar h}\indices{_{\mu\nu,\alpha}^\alpha} - \eta_{\mu\nu} {\bar h}\indices{_{\alpha\beta,}^{\alpha\beta}} + {\bar h}\indices{_{\mu\alpha,}^{\alpha}_{\nu}} + {\bar h}\indices{_{\nu\alpha,}^{\alpha}_{\mu}} \right) = 8\pi T_{\mu\nu}\,,
\end{equation}
where $T_{\mu\nu}$ is the stress-energy tensor. As subsequently discussed in Sec.\ \ref{Supp:Gauge Freedoms in Linearized Gravity}, the latter three terms in Eq.\ \eqref{eq:EinsteinLinear} are ``gauge'' terms, i.e., removable by imposing the subsidiary condition ${\bar h}\indices{^{\mu\alpha}_{,\alpha}} = 0$. This gauge eliminates the derivatives, after which the field equation leads to
\begin{equation}\label{eq:WaveEquationGW}
    {\bar h}\indices{_{\mu\nu,\alpha}^\alpha} = -16\pi T_{\mu\nu}\,, \quad \text{in de Donder gauge:} \quad {\bar h}\indices{^{\mu\alpha}_{,\alpha}} = 0\,.
\end{equation}

\vspace{0.4cm}
\section{Gauge Freedoms in Linearized Gravity}
\label{Supp:Gauge Freedoms in Linearized Gravity}
\vspace{0.4cm}
Focusing on gauge transformations that relate nearly Lorentzian systems, we start by considering the metric in Eq.\ \eqref{eq:PerturbedMetric} in nearly Lorentz coordinates $x^\mu$ and perform a coordinate perturbation $x^{\mu'} = x^\mu + \xi^\mu$, where $\xi^\mu$ is an infinitesimal displacement. Under this perturbation the metric transforms as $g_{\rho'\sigma'} = g_{\mu\nu} \left({\partial x^\mu}/{\partial x^{\rho'}}\right)\left( {\partial x^\nu}/{\partial x^{\sigma'}}\right)$. Substituting Eq.\ \eqref{eq:PerturbedMetric} and differentiating the coordinates we find
$\Lambda\indices{^{\mu'}_{\alpha}} = {\partial x^{\mu'}}/{\partial x^\alpha} = \delta\indices{^\mu_\alpha} + \xi\indices{^\mu_{,\alpha}}$ and $\Lambda\indices{^{\alpha}_{\mu'}} = {\partial x^\alpha}/{\partial x^{\mu'}} = \delta\indices{^\alpha_\mu} - \xi\indices{^\alpha_{,\mu}} + \mathcal{O}\left(\xi^2\right)$, leading to $g_{\rho'\sigma'} = \left(\eta_{\mu\nu} + h_{\mu\nu}\right) \left( \delta\indices{^\mu_\rho} - \xi\indices{^\mu_{,\rho}} \right) \left( \delta\indices{^\nu_\sigma} - \xi\indices{^\nu_{,\sigma}} \right) + \mathcal{O}\left(\xi^2\right)$, or, equivalently, $g_{\rho'\sigma'} = \eta_{\rho\sigma} - \xi\indices{_{\rho,\sigma}} - \xi\indices{_{\sigma,\rho}} + h_{\rho\sigma} + \mathcal{O}\left(\xi^2\right)$. Thus, to first order in $\xi^\mu$, the transformed metric in the new coordinates $x^{\mu'}$ is $g_{\rho'\sigma'} \approx \eta_{\rho\sigma} + h_{\rho\sigma} - \xi\indices{_{\rho,\sigma}} - \xi\indices{_{\sigma,\rho}}$. This relates the metric perturbation in the new coordinates, ${h'}_{\mu\nu}$, to the perturbation in the original coordinates, $h_{\mu\nu}$, via the relationship
\begin{equation}\label{eq:CoordinateChange}
    {h'}_{\mu\nu} = h_{\mu\nu} - \xi\indices{_{\mu,\nu}} - \xi\indices{_{\nu,\mu}}\,.
\end{equation}

Since the components of the Riemann tensor $R\indices{^\alpha_{\beta\mu\nu}}$ are constructed from the metric and its derivatives [cf.\ Eq.\ \eqref{eq:LinearRiemann}], and we are perturbing the coordinates to first order, the Riemann tensor remains invariant to first order in $\xi^\mu$. Consequently the Einstein field equation is unaffected. Starting now with a perturbation $h_{\mu\nu}$ that satisfies Eq.\ \eqref{eq:EinsteinLinear}, the transformed perturbation ${h'}_{\mu\nu}$ also satisfies the de Donder gauge condition, provided that $\xi^\mu$ fulfills the \emph{inhomogeneous} wave equation $ \bar{h}\indices{^{\mu\nu}_{,\nu}} = \xi\indices{^{\mu,\nu}_\nu}$ (see \cite[Exer.\ 18.2]{Misner1973} and \cite[\S 8.36]{Schutz1985}). Thereafter an additional gauge transformation can be applied to ${h'}_{\mu\nu}$, yielding a new perturbation $ (h{''})_{\mu\nu} = h'_{\mu\nu} - \xi'_{\mu,\nu} - \xi'_{\nu,\mu}$. The de Donder condition is preserved, i.e., $(h{''})\indices{^{\mu\nu}_{,\nu}} = 0$, provided that the new gauge choices $\xi'^\mu$ satisfy the \emph{homogeneous} wave equation $ \xi{'}\indices{^{\mu,\nu}_{\nu}} = 0$. 

\vspace{0.4cm}
\section{Gravitational Wave Propagation}
\label{Supp:Gravitational Wave Propagation}
\vspace{0.4cm}
In matter-free spacetime the general solution to the homogeneous Eq.\ \eqref{eq:WaveEquationGW} for a plane GW propagating along the $z$-axis takes the form of the main text, i.e.,
\begin{equation}\label{eq:GWAnsatz}
{\bar h}^{\mu\nu}=\frac{1}{2}\mathcal{H}^{\mu\nu}e^{iK_\rho x^\rho}+\text{c.c.}\,,
\end{equation}
where all the symbols have their usual meaning. Substituting this ansatz into Eq.\ \eqref{eq:WaveEquationGW} yields the condition $K_\mu K^\mu = 0$, which corresponds to the dispersion relation $\omega = k$ showing that {\color{black} non-dispersive} GWs propagate at the phase velocity of light in vacuum (luminal regime). Imposing the de Donder gauge further reveals that $\mathcal{H}_{\mu\nu} K^\mu = 0$, i.e., that the wave’s amplitude is orthogonal to its wavevector.

Additional constraints can be applied to the amplitude $\mathcal{H}_{\mu\nu}$. Indeed, considering a solution $\xi_{\mu\nu} = \mathcal{\bar{W}}_{\mu\nu} e^{iK_\rho x^\rho} + \text{c.c.}$, where $\mathcal{\bar{W}}_{\mu\nu}$ is constant, Eq.\ \eqref{eq:CoordinateChange} implies that \cite{Schutz1985} ${\bar{h}'}_{\mu\nu} = \bar{h}_{\mu\nu} - \xi\indices{_{\mu,\nu}} - \xi\indices{_{\nu,\mu}} + \eta_{\mu\nu} \xi\indices{^{\alpha}_{,\alpha}}$. Omitting the exponential factor gives $\mathcal{H}^{'}_{\mu\nu} = \mathcal{H}_{\mu\nu} - i\mathcal{\bar{W}}_{\mu} K_{\nu} - i\mathcal{\bar{W}}_{\nu} K_{\mu} + i\eta_{\mu\nu} \mathcal{\bar{W}}^{\alpha} K_{\alpha}$, whereby it can be shown that $\mathcal{\bar{W}}_{\mu\nu}$ may be chosen to enforce two additional conditions:
\begin{equation}\label{eq:TTGauge}
    \mathcal{H}\indices{^{\mu}_{\mu}} = 0 \quad \text{(traceless)} \quad \text{and} \quad \mathcal{H}_{\mu\nu} U^\nu = 0,
\end{equation}
where ${\bf U}$ is a fixed 4-velocity. These conditions along with $\mathcal{H}_{\mu\nu} K^\mu = 0$ define the so-called transverse-traceless (TT) gauge.

The TT gauge reduces the number of physical degrees of freedom of the symmetric rank-2 tensor $\bar{h}_{\mu\nu}$ from 10 to 2. Taking the 4-velocity as the time basis vector $U^\nu = \delta\indices{^{\nu}_{0}}$ and orienting the wave along the $z$-axis, the only non-zero components of $\bar{h}_{\mu\nu}$ in the TT gauge are $\bar{h}_{xx}$, $\bar{h}_{yy}$, and $\bar{h}_{xy} = \bar{h}_{yx}$. These components are subject to the traceless condition, $\bar{h}_{xx} + \bar{h}_{yy} = 0$, thereby leaving two \emph{independent} polarization states: $ \bar{h}_{xx} = \mathcal{H}_+ e^{iK_\rho x^\rho} + \text{c.c.}$ and $\bar{h}_{xy} = \mathcal{H}_\times e^{iK_\rho x^\rho} + \text{c.c.}$, where $\mathcal{H}_+$ and $\mathcal{H}_\times$ represent the amplitudes of the ``plus'' and ``cross'' polarization states, respectively—see \cite[Fig.\ 9.1]{Schutz1985} and the discussion therein.

\vspace{0.4cm}
\section{Vanishing Ricci Tensor of Gravitational Waves}
\label{Supp:Vanishing Ricci Tensor of Gravitational Waves}
\vspace{0.4cm}
Under the TT gauge we have ${\bar{h}}_{\mu\nu} = h_{\mu\nu}$ since the trace of the perturbation vanishes, i.e., $h = \bar{h} = h\indices{_\alpha^\alpha} = 0$ (N.B.—this explains the omission of the bar symbol from $h_{\mu\nu}$ in the main text). Substituting the general form of a GW, as given in Eq.\ \eqref{eq:GWAnsatz}, into Eq.\ \eqref{eq:LinearRicciTensor} we obtain
\begin{equation}\label{eq:RicciTensorVanishing}
    R_{\mu\nu} = \frac{1}{2} \left( -K_\mu K_\alpha h\indices{_\nu^\alpha} - K_\nu K_\alpha h\indices{_\mu^\alpha} + K_\alpha K^\alpha h\indices{_{\mu\nu}} - h\indices{_\alpha^\alpha_{,\mu\nu}} \right)\,.
\end{equation}
Here the first two terms are zero on account of transversality $h\indices{_\nu^\alpha}K_\alpha = 0$. The third
term is zero on account of $K_\alpha$ being null {\color{black}(no dispersion)}, and the fourth term is zero on account of the traceless condition of Eq.\ \eqref{eq:TTGauge}. Thus the Ricci tensor for GWs identically vanishes.

\vspace{0.4cm}
\section{Electromagnetic Waves in Curved Spacetime}
\label{Supp:Electromagnetic Waves in Curved Spacetime}
\vspace{0.4cm}
In the absence of sources the electromagnetic field in curved spacetime is characterized by the wave equation for the 4-vector potential $A^\mu$, namely \cite[Eq.\ (22.19b')]{Misner1973}
\begin{equation}\label{eq:wave_eq_curved}
-A\indices{^{\alpha;\mu}_{;\mu}} + A\indices{^\mu_{;\mu}^{;\alpha}}+R\indices{^\alpha_\mu}A^\mu=0\,.
\end{equation}
Thence, under the Lorenz gauge, Eq.~\eqref{eq:wave_eq_curved} simplifies to $A\indices{^{\alpha;\mu}_{;\mu}}=R\indices{^\alpha_{\mu}}A^\mu$. Moreover, as derived in Sec.\ \ref{Supp:Vanishing Ricci Tensor of Gravitational Waves}, the Ricci tensor is zero for GWs under the TT gauge, thus reducing Eq.\ \eqref{eq:wave_eq_curved} to
\begin{equation}\label{eq:WaveEquation}
A\indices{^{\alpha;\mu}_{;\mu}}=0\,, \quad \text{in Lorenz gauge:} \quad A\indices{^{\mu}_{;\mu}}=0\,.
\end{equation}
Subsequently the covariant derivatives in Eq.\ \eqref{eq:WaveEquation} can be expanded as
\begin{eqnarray}\label{eq:WaveEquationExpanded}
0&=&\left(A^{\alpha,\mu}+\eta^{\mu\lambda}\Gamma\indices{^\alpha_{\nu\lambda}}A^\nu\right)_{;\mu}\nonumber\\
&=&
A\indices{^{\alpha,\mu}_{,\mu}}+\eta^{\mu\lambda}\Gamma\indices{^\alpha_{\nu\lambda,\mu}}A^\nu
+\Gamma\indices{^\alpha_{\nu\lambda}}A^{\nu,\lambda}
+\left(A^{\beta,\mu}+\eta^{\mu\lambda}\Gamma\indices{^\beta_{\nu\lambda}}A^\nu\right)\Gamma\indices{^\alpha_{\beta\mu}}\,.
\end{eqnarray}

\par Under the linearized regime outlined in Sec.\ \ref{Supp:Linearized Gravity}, the Christoffel symbols reduce to terms that are first-order in the perturbation $h_{\mu\nu}$ [cf.\ Eq.\ \eqref{eq:LinerChristoffelSymbols}]. Since the Christoffel symbols are proportional to the small perturbation $h_{\mu\nu}$, the term quadratic in $\Gamma$ will be of order $\mathcal{O}\left(h^2\right)$ and can therefore be neglected. This indeed simplifies Eq.\ \eqref{eq:WaveEquationExpanded} to Eq.\ \eqref{eq:WaveEquationFinal} in the main text. In this weak-field limit the TT gauge reduces the Lorenz gauge to $A^{\mu}{}_{,\mu} = 0$, thereby ensuring compatibility: the former constrains the coordinates for $h_{\mu\nu}$, whereas the latter remains metric-independent (i.e., comma goes to semicolon and \emph{thence} back to comma).

We may now consider the metric perturbation of Eq.\ \eqref{eq:GWAnsatz}, which corresponds to a solution of the homogeneous field equation [cf.\ Eq.\ \eqref{eq:WaveEquationGW}], and compute the following:
    \begin{eqnarray*}
    \Gamma\indices{^{\alpha}_{\nu\lambda}} &=& \frac{1}{2}\eta^{\alpha\beta}\left({ h}_{\beta\lambda,\nu}+{ h}_{\nu\beta,\lambda}-{ h}_{\nu\lambda,\beta}\right) = \frac{i}{4}\eta^{\alpha\beta}\Delta\Gamma\indices{_{\beta\nu\lambda}}e^{iK_\gamma x^\gamma}+\text{c.c.}\,,
        \\
        \Gamma\indices{^\alpha_{\nu\lambda,\mu}} &=& \frac{1}{2}\eta^{\alpha\beta}\left({ h}_{\beta\lambda,\nu\mu}+{ h}_{\nu\beta,\lambda\mu}-{ h}_{\nu\lambda,\beta\mu}\right) =\frac{1}{4}\eta^{\alpha\beta}\Delta\Gamma\indices{_{\beta\nu\lambda\mu}}e^{iK_\gamma x^\gamma}+\text{c.c.}
    \end{eqnarray*}
Here, and indeed in the main text, we have defined the auxiliary connection-like objects $\Delta\Gamma\indices{_{\beta\nu\lambda}}=\mathcal{H}_{\beta\lambda}K_\nu+\mathcal{H}_{\nu\beta}K_\lambda-\mathcal{H}_{\nu\lambda}K_\beta$ and $ \Delta\Gamma\indices{_{\beta\nu\lambda\mu}}= \mathcal{H}_{\nu\lambda}K_\beta K_\mu-\mathcal{H}_{\beta\lambda}K_\nu K_\mu-\mathcal{H}_{\nu\beta}K_\lambda K_\mu$.

\vspace{0.4cm}
\section{Archetypal Coupled-Wave Equations}
\label{Supp:Archetypal Coupled-Wave Equations}
\vspace{0.4cm}
As stated in the main text, the total EMW potential is expressed as 
\begin{subequations}
\begin{equation}\label{eq:ExpressionsA1}
    A^\alpha =\sum_j{\cal A }_{(j)}^\alpha  e^{i k_{(j)\mu} x^\mu}\,, 
\end{equation}
so that the remaining terms appearing in Eq.\ \eqref{eq:WaveEquationFinal} are computed as
    \begin{align}
    &A^{\nu,\lambda} = \sum_j \left( {\cal A }_{(j)}^{\nu,\lambda} + i k_{(j)}^\lambda {\cal A }_{(j)}^\nu \right) e^{i k_{(j)\mu} x^\mu} \,, \label{eq:ExpressionsA2}
    \\
&A\indices{^{\alpha,\mu}_{,\mu}} = \sum_j \left( {\cal A }_{(j)}^{\alpha,\mu}{_{,\mu}} 
+ 2ik_{(j)}^{\mu} {\cal A }^{\alpha}_{(j),\mu}
- k_{(j)}^\mu k_{(j)\mu}{\cal A }_{(j)}^\alpha \right) e^{ik_{(j)\nu}x^\nu} \approx \sum_j \left(2ik_{(j)}^{\mu} {\cal A }^{\alpha}_{(j),\mu}
{\color{black}- k_{(j)}^\mu k_{(j)\mu}{\cal A }_{(j)}^\alpha} \right) e^{k_{(j)\nu}x^\nu}\,. \label{eq:ExpressionsA3}
\end{align}
\end{subequations}
We note that the simplification of Eq.\ \eqref{eq:ExpressionsA3} stems from the slowly varying envelope approximation. {\color{black} For EMWs propagating in a plasma, $k_{(j)}$ is no longer null: $k_{(j)}^\mu k_{(j)\mu} = \omega_{(j)}^2(n^2 - 1) \approx -2\delta n\, \omega_{(j)}^2<0$. However, by introducing a constitutive tensor $\chi\indices{^{\alpha\beta\mu\nu}}$, the mode condition $\chi\indices{^{\alpha\beta\mu\nu}}k_\mu k_\beta = 0$ of \cite[Eq.\ (6.35)]{Post1997} eliminates the rightmost term in Eq.\ \eqref{eq:ExpressionsA3}.} Inserting Eqs.\ \eqref{eq:ExpressionsA1}, \eqref{eq:ExpressionsA2}, and \eqref{eq:ExpressionsA3} into Eq.\ \eqref{eq:WaveEquationFinal} leads to Eq.\ \eqref{eq:cwtlong} in the main text,
\begin{align}\label{eq:cwtlong sup}
             &\sum_j\left[2i k_{(j)}^{\mu} {\cal A }^{\alpha}_{(j),\mu} +i\frac{\eta^{\alpha\beta}}{2}\left(\Delta\Gamma\indices{_{\beta\nu\lambda}}e^{iK_\rho x^\rho} - \Delta\Gamma\indices{^{*}_{\beta\nu\lambda}}e^{-iK_\rho x^\rho} \right){\cal A }_{(j)}^{\nu,\lambda}\right]e^{i k_{(j)\gamma}x^\gamma} = \nonumber
        \\
        &\sum_j\left[\frac{\eta^{\alpha\beta}}{2}\left(\Delta\Gamma\indices{_{\beta\nu\lambda}}e^{iK_\rho x^\rho} -\Delta\Gamma\indices{^{*}_{\beta\nu\lambda}}e^{-iK_\rho x^\rho}\right)k_{(j)}^\lambda-\frac{\eta^{\mu\lambda}\eta^{\alpha\beta}}{4}\left(\Delta\Gamma\indices{_{\beta\nu\lambda\mu}}e^{iK_\rho x^\rho}+\Delta\Gamma\indices{^{*}_{\beta\nu\lambda\mu}} e^{-iK_\rho x^\rho}\right) \right]{\cal A }_{(j)}^\nu e^{ik_{(j)\mu} x^\mu}\,.
    \end{align}

Upon inspection of Eq.\ \eqref{eq:cwtlong sup} it becomes evident that potentially synchronous terms will oscillate at the frequencies given by Eq.\ \eqref{eq:phase-match}. Therefrom we may derive the coupled-wave equations for ${\cal A}_{(0)}$ by writing all terms that match $k_{(0)}$ as
\begin{subequations}
    \begin{align}\label{eq:cwe A0}
             &2i k_{(0)}^{\mu} {\cal A }^{\alpha}_{(0),\mu} +i\frac{\eta^{\alpha\beta}}{2}\left(\Delta\Gamma\indices{_{\beta\nu\lambda}}{\cal A }_{(-1)}^{\nu,\lambda} - \Delta\Gamma\indices{^{*}_{\beta\nu\lambda}}{\cal A }_{(1)}^{\nu,\lambda} \right) =\nonumber
             \\
        & \frac{\eta^{\alpha\beta}}{2}\left(k_{(-1)}^\lambda\Delta\Gamma\indices{_{\beta\nu\lambda}}{\cal A }_{(-1)}^\nu -k_{(1)}^\lambda \Delta\Gamma\indices{^{*}_{\beta\nu\lambda}}{\cal A }_{(1)}^\nu\right)-\frac{\eta^{\mu\lambda}\eta^{\alpha\beta}}{4}\left(\Delta\Gamma\indices{_{\beta\nu\lambda\mu}}{\cal A }_{(-1)}^\nu+\Delta\Gamma\indices{^{*}_{\beta\nu\lambda\mu}} {\cal A }_{(1)}^\nu\right)\,.
    \end{align} 
\noindent Thereupon the terms that match $k_{(1)}$ are written as
    \begin{align}\label{eq:cwe A1}
             &2i k_{(1)}^{\mu} {\cal A }^{\alpha}_{(1),\mu} +i\frac{\eta^{\alpha\beta}}{2}\Delta\Gamma\indices{_{\beta\nu\lambda}} {\cal A }_{(0)}^{\nu,\lambda}  = \frac{\eta^{\alpha\beta}}{2}k_{(0)}^\lambda\Delta\Gamma\indices{_{\beta\nu\lambda}} {\cal A }_{(0)}^\nu  
        -\frac{\eta^{\mu\lambda}\eta^{\alpha\beta}}{4}\Delta\Gamma\indices{_{\beta\nu\lambda\mu}} {\cal A }_{(0)}^\nu \,,
    \end{align}

   \noindent  while those that match $k_{(-1)}$ as
    \begin{align}\label{eq:cwe Am1}
             2i k_{(-1)}^{\mu} {\cal A }^{\alpha}_{(-1),\mu} - i\frac{\eta^{\alpha\beta}}{2}\Delta\Gamma\indices{^{*}_{\beta\nu\lambda}} {\cal A }_{(0)}^{\nu,\lambda}  &=-\frac{\eta^{\alpha\beta}}{2}k_{(0)}^\lambda\Delta\Gamma\indices{^{*}_{\beta\nu\lambda}} {\cal A }_{(0)}^\nu  
        -\frac{\eta^{\mu\lambda}\eta^{\alpha\beta}}{4}\Delta\Gamma\indices{^{*}_{\beta\nu\lambda\mu}} {\cal A }_{(0)}^\nu \,.
    \end{align}
\end{subequations}

\par As a final step, Eqs.\ \eqref{eq:cwe A0}, \eqref{eq:cwe A1}, and \eqref{eq:cwe Am1} may be concisely recast as per Eqs.\ (\ref{eq:cwe A0compact}), (\ref{eq:cwe A1compact}), and (\ref{eq:cwe Am1compact}) in the main text, i.e., as
    \begin{subequations}
        \begin{align}\label{eq:cwe A0compact sup}
             k_{(0)}^{\mu} {\cal A }^{\alpha}_{(0),\mu} + C\indices{^\alpha_\nu^\mu}{\cal A }^{\nu}_{(-1),\mu} - \left(C\indices{^\alpha_\nu^\mu}\right)^*{\cal A }^{\nu}_{(1),\mu} &= P\indices{^\alpha_\nu}{\cal A}^\nu_{(-1)}+Q\indices{^\alpha_\nu}{\cal A}^\nu_{(1)}\,,
             \\
             k^\mu_{(1)}{\cal A}^\alpha_{(1),\mu} + C\indices{^\alpha_\nu^\mu}{\cal A}^\nu_{(0),\mu} &=M\indices{^\alpha_\nu}{\cal A}^\nu_{(0)}+iN\indices{^\alpha_\nu}{\cal A}^\nu_{(0)}\,,
                \label{eq:cwe A1compact sup}
                \\
k^\mu_{(-1)}{\cal A}^\alpha_{(-1),\mu} - \left(C\indices{^\alpha_\nu^\mu}\right)^*{\cal A}^\nu_{(0),\mu} &= \left(M\indices{^\alpha_\nu}\right)^*{\cal A}^\nu_{(0)} +i \left(N\indices{^\alpha_\nu}\right)^*{\cal A}^\nu_{(0)}\,,
                \label{eq:cwe Am1compact sup}  
    \end{align}
\end{subequations}
wherein each tensor appearing above is defined as per the main text:
\begin{align*}
    &C\indices{^{\alpha}_{\nu}^{\mu}} = \frac{\eta^{\alpha\beta}}{4} \Delta\Gamma\indices{_{\beta\nu\lambda}}\eta^{\lambda\mu}\,, \  M\indices{^\alpha_\nu} = -i\frac{\eta^{\alpha\beta}}{4}k_{(0)}^\lambda\Delta\Gamma\indices{_{\beta\nu\lambda}}\,, \ N\indices{^\alpha_\nu} = \frac{\eta^{\mu\lambda}\eta^{\alpha\beta}}{8}\Delta\Gamma_{\beta\nu\lambda\mu}\,,
    \\
    &Q\indices{^\alpha_\nu} = \frac{i}{4}\left(\eta^{\alpha\beta}k_{(1)}^\lambda\Delta\Gamma\indices{^*_{\beta\nu\lambda}}+\frac{\eta^{\mu\lambda}\eta^{\alpha\beta}}{2} \Delta\Gamma\indices{^*_{\beta\nu\lambda\mu}}\right)\,, \  P\indices{^\alpha_\nu} = -\frac{i}{4}\left(\eta^{\alpha\beta}k^\lambda_{(-1)}\Delta\Gamma_{\beta\nu\lambda}-\frac{\eta^{\mu\lambda}\eta^{\alpha\beta}}{2}\Delta\Gamma_{\beta\nu\lambda\mu}\right)\,.
\end{align*}
    
\vspace{0.4cm}
\section{Polarization-Specific Coupled-Wave Equations}
\label{Polarization-Specific Coupled-Wave Equations}
\vspace{0.4cm}
Although Eqs.\ \eqref{eq:cwe A0compact sup}, \eqref{eq:cwe A1compact sup}, and \eqref{eq:cwe Am1compact sup} [or indeed Eqs.\ (\ref{eq:cwe A0compact}), (\ref{eq:cwe A1compact}), and (\ref{eq:cwe Am1compact}) of the main text] constitute a key result of this work, it is instructive to specify the polarizations of the EMWs and of the GW to gain further insights. Henceforth the direction of co-propagation is assumed to lie along the $z$-axis. The transverse components of the incident electromagnetic potential may be written as {\color{black}$A_{x} = \mathcal{A}_{x}(t,z) e^{i\omega_{(0)} (nz - t)}$ and $A_{y} = \mathcal{A}_{y}(t,z) e^{i\omega_{(0)} (nz - t)}$}, i.e., $k_{(0)\mu} = \omega_{(0)}(-1, 0, 0, n)$. Recalling that the Faraday tensor is given by $F_{\mu\nu} = A_{\nu,\mu} - A_{\mu,\nu}$, we see that these generate non-zero electric and magnetic field components, viz., {\color{black}$E_x = (i\omega_{(0)} \mathcal{A}_x - \partial_t \mathcal{A}_x) e^{i\omega_{(0)} (nz - t)}$, $E_y = (i\omega_{(0)} \mathcal{A}_y - \partial_t \mathcal{A}_y) e^{i\omega_{(0)} (nz - t)}$, $B_x = -(\partial_z \mathcal{A}_y + i n \omega_{(0)} \mathcal{A}_y) e^{i\omega_{(0)} (nz - t)}$, and $B_y = (\partial_z \mathcal{A}_x + i n \omega_{(0)} \mathcal{A}_x) e^{i\omega_{(0)} (nz - t)}$}; here $\partial_t \equiv \partial / \partial t$.

The Lorenz gauge condition thus reduces to $A\indices{^x_{,x}}+\Gamma\indices{^\mu_{x\mu}}A^x +A\indices{^y_{,y}}+\Gamma\indices{^\mu_{y\mu}}A^y=0$, whereby the first and third terms are clearly zero. Regarding the second term, $\Gamma\indices{^\mu_{x\mu}} = \left({\eta^{\mu\nu}}/{2}\right)\left(h_{\nu\mu,x}+h_{x\nu,\mu}-h_{x\mu,\nu}\right)$, the last two terms cancel, and the first is zero as there is no transverse variation for a plane GW propagating along $z$. Hence $\Gamma\indices{^\mu_{x\mu}}=0$ and similarly $\Gamma\indices{^\mu_{y\mu}}=0$ so that the Lorenz gauge is $A\indices{^\mu_{,\mu}}=0$—see also the discussion on the ``comma-goes-to-semicolon-and-thence-back-to-comma'' rule in Sec.\ \ref{Supp:Electromagnetic Waves in Curved Spacetime}.

Having specified the polarizations, we can now compute the non-zero independent components of the tensors appearing in Eqs.\ \eqref{eq:cwe A0compact sup}, \eqref{eq:cwe A1compact sup}, and \eqref{eq:cwe Am1compact sup}. Apart from  $N\indices{^\alpha_\nu}$ which does not have any non-zero components for the specified polarization, the rest read:
\vspace{0.5cm}
\par \textbullet \quad \underline{{\bf Tensor $ C\indices{^{\alpha}_{\nu}^{\mu}}$}} 
\begin{align*}
    &C\indices{^{0}_{1}^{1}} = -\frac{1}{4} \mathcal{H}_+ \Omega\,, \ C\indices{^{0}_{1}^{2}} = C\indices{^{0}_{2}^{1}}= -\frac{1}{4} \mathcal{H}_\times \Omega\,, \ C\indices{^{0}_{2}^{2}} = \frac{1}{4} \mathcal{H}_+ \Omega\,,
    \\
    &C\indices{^{1}_{0}^{1}} = -C\indices{^{1}_{1}^{0}} = -\frac{1}{4} \mathcal{H}_+ \Omega\,, \ C\indices{^{1}_{0}^{2}} = -C\indices{^{1}_{2}^{0}} = -\frac{1}{4} \mathcal{H}_\times \Omega,  
        C\indices{^{1}_{1}^{3}} = C\indices{^{1}_{3}^{1}} = \frac{1}{4} \mathcal{H}_+ \Omega\,, \ C\indices{^{1}_{2}^{3}} = C\indices{^{1}_{3}^{2}} = \frac{1}{4} \mathcal{H}_\times \Omega, 
    \\
    &C\indices{^{2}_{0}^{1}} = -C\indices{^{2}_{1}^{0}} = -\frac{1}{4} \mathcal{H}_\times \Omega\,, \ C\indices{^{2}_{0}^{2}} = -C\indices{^{2}_{2}^{0}} = \frac{1}{4} \mathcal{H}_+ \Omega,  
        C\indices{^{2}_{1}^{3}} = C\indices{^{2}_{3}^{1}} = \frac{1}{4} \mathcal{H}_\times \Omega\,, \ C\indices{^{2}_{2}^{3}} = C\indices{^{2}_{3}^{2}} = -\frac{1}{4} \mathcal{H}_+ \Omega,
    \\
    &C\indices{^{3}_{1}^{1}} = -\frac{\Omega}{4} \mathcal{H}_+\,, \ C\indices{^{3}_{1}^{2}} = C\indices{^{3}_{2}^{1}}  = -\frac{\Omega}{4} \mathcal{H}_\times\,, \ C\indices{^{3}_{2}^{2}} = \frac{\Omega}{4} \mathcal{H}_+.
\end{align*}
    
\par \textbullet \quad \underline{{\bf Tensor $M\indices{^\alpha_\nu}$}}
\begin{equation*}
M\indices{^1_1} = {\color{black}i\frac{\delta n}{4}} \mathcal{H}_+ \Omega \omega_{(0)}, \quad
M\indices{^1_2} = {\color{black}i\frac{\delta n}{4}} \mathcal{H}_\times \Omega \omega_{(0)}, \quad
M\indices{^2_1} = {\color{black}i\frac{\delta n}{4}} \mathcal{H}_\times \Omega \omega_{(0)}, \quad
M\indices{^2_2} = {\color{black}-i\frac{\delta n}{4}} \mathcal{H}_+ \Omega \omega_{(0)}.
\end{equation*}

\par \textbullet \quad \underline{{\bf Tensor $Q\indices{^\alpha_\nu}$}}
 \begin{align*}
Q\indices{^1_1} = {\color{black}-i\frac{\delta n}{4}} \mathcal{H}_+^* \Omega \omega_{(1)}, \quad
Q\indices{^1_2} = {\color{black}-i\frac{\delta n}{4}} \mathcal{H}_\times^* \Omega \omega_{(1)}, \quad
Q\indices{^2_1} = {\color{black}-i\frac{\delta n}{4}} \mathcal{H}_\times^* \Omega \omega_{(1)}, \quad
Q\indices{^2_2} = {\color{black}i\frac{\delta n}{4}} \mathcal{H}_+^* \Omega \omega_{(1)}.
\end{align*}

\par \textbullet \quad \underline{{\bf Tensor $P\indices{^\alpha_\nu}$}}
 \begin{align*}
P\indices{^1_1} = {\color{black}i\frac{\delta n}{4}} \mathcal{H}_+ \Omega \omega_{(-1)}, \quad
P\indices{^1_2} = {\color{black}i\frac{\delta n}{4}} \mathcal{H}_\times \Omega \omega_{(-1)}, \quad
P\indices{^2_1} = {\color{black}i\frac{\delta n}{4}} \mathcal{H}_\times \Omega \omega_{(-1)}, \quad
P\indices{^2_2} = {\color{black}-i\frac{\delta n}{4}} \mathcal{H}_+ \Omega \omega_{(-1)}.
\end{align*}

\par Setting $\alpha\equiv x$ in Eqs.\ \eqref{eq:cwe A0compact sup}, \eqref{eq:cwe A1compact sup}, and \eqref{eq:cwe Am1compact sup}, and inserting the relevant coefficients calculated above, we obtain, respectively,
\begin{eqnarray*}
    &\omega_{(0)} {\rm d}_u{\cal A}^x_{(0)} -\frac{\Omega}{4}{\rm d}_u\left({\cal H}_+{\cal A}^x_{(-1)}+{\cal H}_\times{\cal A}^y_{(-1)}\right)+\frac{\Omega}{4}{\rm d}_u\left({\cal H}_+^*{\cal A}^x_{(1)}+{\cal H}_\times^*{\cal A}^y_{(1)}\right) 
    \\
    &={\color{black}i\frac{\delta n}{4}}\Omega\omega_{(-1)}\left({\cal H}_+{\cal A}^x_{(-1)}+{\cal H}_\times{\cal A}^y_{(-1)}\right){\color{black}-i\frac{\delta n}{4}}\Omega\omega_{(1)}\left({\cal H}_+^*{\cal A}^x_{(1)}+{\cal H}_\times^*{\cal A}^y_{(1)}\right)\,,
    \\
    &\omega_{(1)}{\rm d}_u{\cal A}^x_{(1)} +\frac{\Omega}{4}{\rm d}_u\left({\cal H}_+{\cal A}^x_{(0)} + {\cal H}_\times{\cal A}^y_{(0)}\right) = {\color{black}i\frac{\delta n}{4}}\Omega\omega_{(0)}\left({\cal H}_+{\cal A}^x_{(0)}+{\cal H}_\times{\cal A}^y_{(0)}\right)\,,
    \\
     &\omega_{(-1)}{\rm d}_u{\cal A}^x_{(-1)} +\frac{\Omega}{4}{\rm d}_u\left({\cal H}_+^*{\cal A}^x_{(0)} + {\cal H}_\times^*{\cal A}^y_{(0)}\right)={\color{black}-i\frac{\delta n}{4}}\Omega\omega_{(0)}\left({\cal H}_+^*{\cal A}^x_{(0)}+{\cal H}_\times^*{\cal A}^y_{(0)}\right)\,.
\end{eqnarray*}
Repeating the same process for $\alpha\equiv y$ yields, respectively,
\begin{eqnarray*}
    &\omega_{(0)} {\rm d}_u{\cal A}^y_{(0)} -\frac{\Omega}{4}{\rm d}_u\left( {\cal H}_\times{\cal A}^x_{(-1)}-{\cal H}_+{\cal A}^y_{(-1)}\right)+\frac{\Omega}{4}{\rm d}_u\left({\cal H}_\times^*{\cal A}^x_{(1)}-{\cal H}_+^*{\cal A}^y_{(1)} \right) 
    \\
    &={\color{black}i\frac{\delta n}{4}}\Omega\omega_{(-1)}\left({\cal H}_\times{\cal A}^x_{(-1)}-{\cal H}_+{\cal A}^y_{(-1)}\right){\color{black}-i\frac{\delta n}{4}}\Omega\omega_{(1)}\left({\cal H}_\times^*{\cal A}^x_{(1)}-{\cal H}_+^*{\cal A}^y_{(1)}\right)\,,
    \\
    &\omega_{(1)}{\rm d}_u{\cal A}^y_{(1)} +\frac{\Omega}{4}{\rm d}_u\left({\cal H}_\times{\cal A}^x_{(0)}-{\cal H}_+{\cal A}^y_{(0)}\right)= {\color{black}i\frac{\delta n}{4}}\Omega\omega_{(0)}\left({\cal H}_\times{\cal A}^x_{(0)}-{\cal H}_+{\cal A}^y_{(0)}\right)\,,
    \\
    &\omega_{(-1)}{\rm d}_u{\cal A}^y_{(-1)} +\frac{\Omega}{4}{\rm d}_u\left({\cal H}_\times^*{\cal A}^x_{(0)}-{\cal H}_+^*{\cal A}^y_{(0)}\right)= {\color{black}-i\frac{\delta n}{4}}\Omega\omega_{(0)}\left({\cal H}_\times^*{\cal A}^x_{(0)}-{\cal H}_+^*{\cal A}^y_{(0)}\right)\,.
\end{eqnarray*}

\par The equations above may be compactly expressed in a matrix notation as per Eq.\ \eqref{CWEs General} in the main text, namely
\begin{equation}\label{CWEs General Sup}
    \mathbf{L} \cdot {\rm d}_u \mathbf{A} = \mathbf{S} \cdot \mathbf{A}\,,
\end{equation}
where the characteristic matrices are precisely those presented in the main text, which in a not-so-concise representation assume the forms
\begin{equation*}
    \mathbf{L} =
\begin{bmatrix}
\omega_{(-1)} & 0 & \frac{\Omega}{4}{\cal H}_+^* & \frac{\Omega}{4}{\cal H}_\times^* & 0 & 0 \\
0 & \omega_{(-1)} & \frac{\Omega}{4}{\cal H}_\times^* & -\frac{\Omega}{4}{\cal H}_+^* & 0 & 0 \\
-\frac{\Omega}{4}{\cal H}_+ & -\frac{\Omega}{4}{\cal H}_\times & \omega_{(0)} & 0 & \frac{\Omega}{4}{\cal H}_+^* & \frac{\Omega}{4}{\cal H}_\times^* \\
-\frac{\Omega}{4}{\cal H}_\times & \frac{\Omega}{4}{\cal H}_+ & 0 & \omega_{(0)} & \frac{\Omega}{4}{\cal H}_\times^* & -\frac{\Omega}{4}{\cal H}_+^* \\
0 & 0 & \frac{\Omega}{4}{\cal H}_+ & \frac{\Omega}{4}{\cal H}_\times & \omega_{(1)} & 0 \\
0 & 0 & \frac{\Omega}{4}{\cal H}_\times & -\frac{\Omega}{4}{\cal H}_+ & 0 & \omega_{(1)}
\end{bmatrix}
\end{equation*}
and
\begin{equation*}
    \mathbf{S} = i{\color{black}\delta n}\frac{\Omega}{4}
\begin{bmatrix}
0 & 0 & -\omega_{(0)} {\cal H}_+^* & -\omega_{(0)} {\cal H}_\times^* & 0 & 0 \\
0 & 0 & -\omega_{(0)} {\cal H}_\times^* & \omega_{(0)} {\cal H}_+^* & 0 & 0 \\
\omega_{(-1)} {\cal H}_+ & \omega_{(-1)} {\cal H}_\times & 0 & 0 & -\omega_{(1)} {\cal H}_+^* & -\omega_{(1)} {\cal H}_\times^* \\
\omega_{(-1)} {\cal H}_\times & -\omega_{(-1)} {\cal H}_+ & 0 & 0 & -\omega_{(1)} {\cal H}_\times^* & \omega_{(1)} {\cal H}_+^* \\
0 & 0 & \omega_{(0)} {\cal H}_+ & \omega_{(0)} {\cal H}_\times & 0 & 0 \\
0 & 0 & \omega_{(0)} {\cal H}_\times & -\omega_{(0)} {\cal H}_+ & 0 & 0
\end{bmatrix}\,.
\end{equation*}
Evidently, insofar as $\mathbf{L}$ is invertible, we can define $\mathbf{W} = \mathbf{L}^{-1} \cdot \mathbf{S}$ so that the coupled-wave system in Eq.\ \eqref{eq:4dcwequation} in the main text follows directly from Eq.\ \eqref{CWEs General Sup} [or indeed from Eq.\ \eqref{CWEs General}].

Although Eq.\ \eqref{eq:4dcwequation} can be readily solved numerically—this is done to produce Fig.\ 2—key observations provide insights into the coupling strengths. To begin with, phase-matching requires that $\omega_{(1)} = \omega_{(0)} + \Omega$ and $\omega_{(-1)} = \omega_{(0)} - \Omega$. Furthermore, for the parameters provided in the caption of Fig.\ 2,  the diagonal elements of $\mathbf{L}$ vastly outweigh the $\mathcal{O}(h_s)$ ($h_s$ being either $\left|\mathcal{H}_{+}\right|$ or $\left|\mathcal{H}_{\times}\right|$) off-diagonal terms. Consequently Eq.\ \eqref{eq:4dcwequation} reduces to
\begin{equation}\label{CWEsReduced}
   {\rm d}_{u}\begin{bmatrix}
{\cal A}^x_{(-1)} \\
{\cal A}^y_{(-1)} \\
{\cal A}^x_{(0)} \\
{\cal A}^y_{(0)} \\
{\cal A}^x_{(1)} \\
{\cal A}^y_{(1)}
\end{bmatrix} = \begin{bmatrix}
0 & 0 & i\kappa^{x,x}_{-1,0} & i\kappa^{x,y}_{-1,0}  & 0 & 0 \\
0 & 0 & i\kappa^{x,y}_{-1,0} & -i\kappa^{x,x}_{-1,0} & 0 & 0 \\
i\kappa^{x,x}_{0,-1} & i\kappa^{x,y}_{0,-1} & 0 & 0 &  i\kappa^{x,x}_{0,1} & i\kappa^{x,y}_{0,1} \\
i\kappa^{x,y}_{0,-1} & -i\kappa^{x,x}_{0,-1} & 0 & 0 & i\kappa^{x,y}_{0,1} & - i\kappa^{x,x}_{0,1} \\
0 & 0 & i\kappa^{x,x}_{1,0} & i\kappa^{x,y}_{1,0} & 0 & 0 \\
0 & 0 & i\kappa^{x,y}_{1,0} & -i\kappa^{x,x}_{1,0} & 0 & 0
\end{bmatrix}\begin{bmatrix}
{\cal A}^x_{(-1)} \\
{\cal A}^y_{(-1)} \\
{\cal A}^x_{(0)} \\
{\cal A}^y_{(0)} \\
{\cal A}^x_{(1)} \\
{\cal A}^y_{(1)}
\end{bmatrix}\,,
\end{equation}
wherein the identified coupling coefficients read
\begin{align*}
    &\kappa^{x,x}_{-1,0} = {\color{black}\delta n}\frac{\omega_{(0)} \Omega}{{\color{black}4}\omega_{(0)} - {\color{black}4}\Omega}\mathcal{H}_+^*\,, \ \kappa^{x,y}_{-1,0} = {\color{black}\delta n}\frac{\omega_{(0)} \Omega}{{\color{black}4}\omega_{(0)} - {\color{black}4}\Omega}\mathcal{H}_\times^*\,, \ 
    \kappa^{x,x}_{0,-1} = {\color{black}\delta n}\frac{\Omega^2-\omega_{(0)}\Omega}{{\color{black}4}\omega_{(0)}}\mathcal{H}_+\,, \ \kappa^{x,y}_{0,-1} = {\color{black}\delta n}\frac{\Omega^2-\omega_{(0)}\Omega}{{\color{black}4}\omega_{(0)}}\mathcal{H}_\times\,, 
    \\
    &\kappa^{x,x}_{0,1} = {\color{black}\delta n}\frac{\omega_{(0)}\Omega + \Omega^2}{{\color{black}4}\omega_{(0)}}\mathcal{H}_+^*\,, \ \kappa^{x,y}_{0,1} = {\color{black}\delta n}\frac{\omega_{(0)}\Omega + \Omega^2}{{\color{black}4}\omega_{(0)}}\mathcal{H}_\times^*\,, \
    \kappa^{x,x}_{1,0} = -{\color{black}\delta n}\frac{\omega_{(0)}\Omega}{{\color{black}4}\omega_{(0)} + {\color{black}4}\Omega}\mathcal{H}_+\,, \ \kappa^{x,y}_{1,0}=-{\color{black}\delta n}\frac{\omega_{(0)}\Omega}{{\color{black}4}\omega_{(0)} + {\color{black}4}\Omega}\mathcal{H}_\times\,.
\end{align*}

\par Given that the characteristic interaction length is inversely proportional to the coupling strength, the key approximation in Eq.\ \eqref{eq:estimation} of the main text is thereby substantiated. {\color{black}The dispersion parameter, $\delta n$, which quantifies the relative motion between the nearly-luminal impinging EMW and the luminal GW,} along with the ratio $\Lambda/h_s$, emerge as the fundamental parameters governing the interaction length(s).

\begin{center}
    \textit{FINIS}
\end{center}

\end{document}